\documentclass[preprint,3p,twocolumn,number]{elsarticle}

\usepackage{graphicx}
\usepackage{amsmath}
\usepackage{amssymb}
\usepackage{dcolumn}

\usepackage{color,soul}

\journal{Journal Of Colloid and Interface Science}

\begin{document}

\begin{frontmatter}

\title{Morphology of anisotropic chains in a magneto-rheological fluid during aggregation and disaggregation processes.}

\author[a]{P. Dom\'inguez-Garc\'ia}
\ead{pdominguez@fisfun.uned.es}
\author[b]{Sonia Melle}
\author[c]{M.A. Rubio}
\address[a]{Dep. F\'isica de Materiales, UNED, Senda del Rey 9, Madrid 28040, Spain.}
\address[b]{Dep. \'Optica, Universidad Complutense de Madrid, Arcos de Jal\'on s/n 28037 Madrid, Spain.}
\address[c]{Dep. F\'isica Fundamental, UNED, Senda del Rey 9,
Madrid 28040, Spain.}

\begin{abstract}
We study the morphology of the chain-like aggregates formed when a
external constant and uniaxial magnetic field is applied to a
magneto-rheological (MR) fluid. In order to characterize the
conformation of the aggregates, we study the evolution of various
fractal dimensions during aggregation and
disaggregation processes (i.e., when the applied field is switched on and off), 
using video-microscopy and image analysis. Experiments have been performed by varying the values of two external parameters: the magnetic field amplitude and particle concentration. We found that the box-counting dimension, related with how the aggregates occupy the surrounding space, depends on the ratio $R_1/R_0$. During the first stage of the disaggregation process, when the particles are moving by Brownian motion inside the aggregate, Family-Vicsek scaling function is verified.
\end{abstract}


\end{frontmatter}

\section{Introduction.}\label{introduction}

A magneto-rheological (MR) fluid \cite{Rabinow48} is a colloidal dispersion of micron-sized paramagnetic particles in some carrier fluid. Under the action of a constant uniaxial magnetic field, these particles acquire a magnetic dipole moment
that forces them to aggregate into linear chains \cite{Kerr}. When the external field is removed, the inverse processes occur, named disaggregation. The main interest on these fluids lay on applications. In fact, the rheological properties of these fluids vary drastically when a magnetic field is applied. Therefore, MR fluids have been widely used for the design of on-off dispositives for industrial applications \cite{Nakano98}. Magnetic particles are used also for biomedical research \cite{cortisol,tumortherapy}. From an applied-physics point of view, the adequate knowledge of the morphology of the aggregates during aggregation and disaggregation processes is important to characterize the cluster behavior on the different applications.

The morphology of the aggregates using MR fluids has been studied in a few works \cite{Helgesen1988, Martinez-Pedrero2005, Martinez-Pedrero2006, Martinez-PedreroCSA2007}. In some cases, the growth of aggregates shows scale invariance that can be described in terms of fractal dimensions \cite{viskecbook,jullienbook}. Magnetic particles \cite{Helgesen1988,Ding1989,Niklasson1988}, magneto-rheological fluids
\cite{Carrillo2003,Martinez-Pedrero2005, Martinez-Pedrero2006, Martinez-PedreroCSA2007},
organic-solvent-based magnetic fluids \cite{Shen2001} or magnetic liposomes \cite{Licinio2001} have been the object of several experimental investigations. 

Two main classes of optical techniques are widely used in the study of the morphology of the aggregates in colloid science:
light scattering techniques \cite{Bushell2002} and direct imaging techniques (e.g. video-microscopy \cite{Crocker1996}). Light scattering techniques gives a fractal dimension, $D_f$, related to a three-dimensional structure. It can be obtained by adjusting the scattering intensity $I$ with the scattering vector $q$ between the Guinier and the Porod regimes, where $I \propto q^{-D_f} $\cite{Bushell2002} is verified. All of the values relating chain-like
structures of magnetic particles are contained in an interval $D_f = 1.1 - 1.3$ \cite{Licinio2001, Shen2001, Martinez-Pedrero2005, Martinez-Pedrero2006, Martinez-PedreroCSA2007}, reflecting a linear structure of the aggregates, because of the alignment with the external field.  However, imaging techniques give only two-dimensional information of the aggregates. For example, Helgesen \emph{et al.} \cite{Helgesen1988} reported variation of fractal dimension with the magnetic strength for a 2D system of $3.6~\mu$m magnetic particles contained in a cell of $5~\mu$m
thickness. Then, the fractal dimension is calculated using the number of primary particles in the system that are detected using video-microscopy. The relation of the values of $D_f$ and the two dimensional fractal dimensions is not an straightforward matter as has been pointed out in \cite{Lee2004, Maggi2004, Sanchez2005}.

In this work, we focus on the characterization
of the aggregates morphology in a MR fluid in terms of fractal properties using image analysis, with the aim of characterizing the formation and disappearance of the aggregates. 

\section{Materials and Methods}\label{experimental}

\subsection{Experimental Setup and methodology.}

\begin{figure}
\begin{center}
\includegraphics[scale=1]{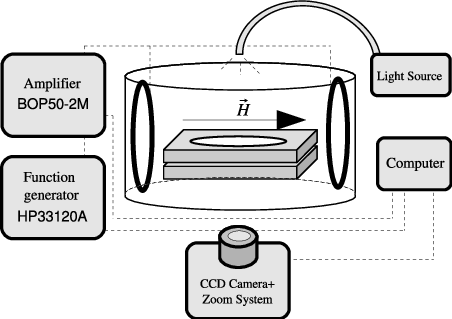}
\end{center}
\caption{\label{fig:setup}Scheme of the video-microscopy setup \cite{Dominguez-Garcia2007}.}
\end{figure}

The magneto-rheological sample consists on a commercial aqueous
suspension of super-paramagnetic particles (Estapor M1-070/60).
These particles are composed of a polystyrene matrix with embedded
magnetite crystals ($Fe_{3}O_{4}$) of small size ($\sim 10$ nm).
Since these iron oxide grains are randomly oriented inside the
micro-particles, the resulting average magnetic moment is
negligible in the absence of an external magnetic field. The
super-paramagnetic particles have a radius $a = 0.48$ $\mu m$ and
a density $\rho \sim 1.85$ g/$cm^{3}$ with a magnetic content of
$54.7\%$ in weight.

When an external magnetic field $\vec{H}$ is applied, these
particles acquire a magnetic dipole moment aligned with the
external field, i.e., $\vec{m} = (4\pi/3))a^{3}\vec{M}$ , with
$\vec{M}=\chi\vec{H}$, where $\vec{m}$ is the magnetic moment of
the particles, $\vec{M}$ is the magnetization of the particle and
$\chi$ the particle magnetic susceptibility, being the magnetic saturation $42$ kA/m (23 emu/gr) \cite{Dominguez-Garcia2007}. The surface of the latex micro-spheres is also functionalized with carboxylic groups. The suspension has been also provided with a
$1$ g/l concentration of a surfactant (sodium dodecil sulfate). The carboxylic groups prevent spontaneous aggregation of the particles, 
while the surfactant facilitates the disaggregation process.

A diagram of the system that generates the magnetic field, the
cell containing the MR fluid and the video-microscopy setup can be
seen on Fig. \ref{fig:setup}. A thermostatic bath keeps the sample temperature constant to $T = 282 \pm 1$ K. Further details of the experimental setup can be consulted on Ref. \cite{Dominguez-Garcia2007}.

We capture images during $5$ minutes without field
and then, a constant uniform magnetic field is switched on. The
field triggers the aggregation process that we register during approximately 5000
seconds. Finally, we switch off the magnetic field and capture
another hour of images during the disaggregation process of the
previously formed chains. The image analysis, data extraction and
statistical calculations have been carried out with image
processing software developed at our lab, based in \emph{ImageJ} \cite{ImageJ2}.

\begin{figure}
\begin{center}
\includegraphics[scale=0.95]{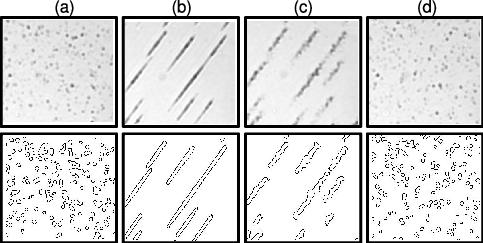}
\end{center} \caption{\label{fig:circ}The four stages of the experiment are (a) free
particles in brownian motion, (b) chains formed after applying the field during 4000 seconds, (c) chains disaggregating when the magnetic field is switched off and (d) free particles again. Upper row: direct images; lower row: cluster contours after image processing.}
\end{figure}

The described methodology used in the experiments captures the four principal stages that can be found on applications using a MR fluid, i.e., suspension without field,
aggregation upon application of the magnetic field, switched off and finally,
return to the initial disordered stage. In Fig. \ref{fig:circ}, we show four examples of images of clusters, in their original form (top) and after being
processed (bottom) on the four stages. 

The control parameters in the experiments are the volume fraction
of the suspension, $\phi$, and the ratio between the magnetic
interaction energy for two particles in contact, $W_{m}$ and the
energy of thermal fluctuations, $k_{B}T$:
\begin{equation}\label{lambda}
\lambda \equiv \frac{W_{m}}{k_{B}T} = \frac{\mu_{0}m^{2}}{16\pi
a^{3}k_{B}T}
\end{equation}
where $\mu_{0}$ is the vacuum magnetic permeability, $k_{B}$ the
Boltzmann constant, and $T$ the temperature.

The volume fraction of particles $\phi$ in the solution is defined
as the fraction of the volume occupied by the particles over the
total volume of the solution. For the purpose of this study, it is
more useful to use an effective surface fraction, which is
calculated by dividing the total area occupied by all the clusters
contained in the image by the image total area. Hereafter,
$\phi_{2D}$ will refer to this effective surface fraction. These
parameters, $\lambda$ and $\phi_{2D}$, can be used to define two
characteristic length scales. The first quantity is the distance,
$R_1$, at which the dipole-dipole interaction energy is equal to
the energy of thermal fluctuations, i.e., $R_1 \equiv
2a\lambda^{1/3}$. The second one is an initial average
inter-particle distance $R_0 \approx 2a\phi_{2D}^{1/2}$. The ratio
of these two length scales allows us to distinguish between
diffusion limited and field driven aggregation processes. If $R_1 < R_0$ at
the time the field is switched on, the aggregation
process should be diffusion limited, whereas if $R_1
> R_0$, the aggregation process should be field driven.

\subsection{Fractal dimension using image analysis.}\label{fractal_methods}

Concerning particle aggregates, the fractal dimension, $D_f$
(also labelled $D_3$ or $D$) is usually calculated by means of the expression $N \sim R_g^{D_{f}}$, where $N$, is its number of particles and $R_g$ its radius of gyration \cite{viskecbook}. However, it is usual in image analysis not to have a reliable experimental setup or methodology for the right extraction of the number of particles per cluster. In our case, we have searched for a reasonable balance between good statistics and spatial resolution. As a result, we do not have enough image definition for detecting the individual particles inside each cluster (see images on Fig. \ref{fig:circ}) and only the contours of the clusters have
been detected.

\begin{figure}
\begin{center}
\includegraphics[scale=1]{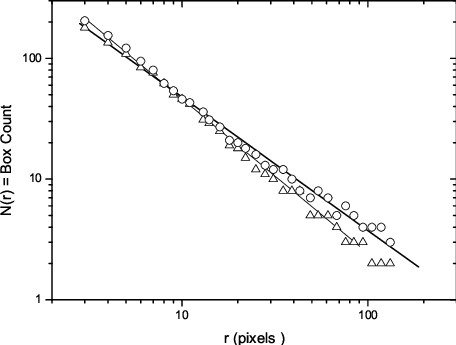}
\end{center} \caption{\label{db_example} Log-log plots of number of counts
versus the box size. We use a chain during aggregation
($t=4989.6$ s under the action of the field) and during disaggregation ($10$ s after the field
has been switched off) for an experiment with $\lambda=$812 and
$\phi_{2D}=$0.106. In the first case we obtain $D_B(2D)=1.11 \pm
0.02$ and $D_B(2D)=1.28 \pm 0.02$ in the second.}
\end{figure}

If we cover a fractal object with a number of boxes $N(r)$ with side $r$, we can obtain its capacity dimension using the following expression \cite{Smith1996}:
\begin{equation}\label{capacity}
D_B=-\lim_{r\rightarrow0}\frac{\log N(r)}{\log r}
\end{equation}
This method gives the so called box-counting dimension or capacity dimension ($D_B$ hereafter). This calculation is very sensitive to the resolution and orientation of the image and the power-law relationship must be verified in a reasonable range of length scales to be able to state that a fractal structure
is present \cite{Halley2004}. This method is applicable both to single objects (single cluster) or to a spatial distribution of objects.

In our experiments, we calculate the capacity dimension $D_B(2D)$ using several
clusters in some images during aggregation and disaggregation. For an adequate calculation of this dimension, we manually extract well-defined representative clusters in each image and calculate individually the box-counting dimension.  This methodology is only valid when we study objects with enough size to fit a linear regression in an acceptable range. For making a correct $D_B(2D)$ calculation we need a chain almost 8 microns-long. The clusters on a single time have a very similar aspect in terms of their shape, as can be seen on Fig. \ref{fig:circ}. Therefore, we choose eight to ten long chains for making an average of $D_B(2D)$ for the corresponding analyzed time. Next, we make an average value of $D_B(2D)$ for each analyzed time. Besides, because of the image resolution, we cannot use this method in the stages in which the number of free particles is predominant. For small objects, like free particles, we do not have enough precision -or number of pixels- to clearly observe the border of the particles, so we cannot suitably obtain their corresponding fractal dimension. An example can be seen on Fig. \ref{db_example} where we  calculate the two-dimensional box-counting dimension, that we name as $D_B(2D)$, for a chain during aggregation, $5000$ s after the field was switch on, and for disaggregation $10$ s after the field has been turned off. In the first case, a capacity dimension $D_B(2D)=1.11 \pm 0.02$ is obtained, whereas in the second, we obtain $D_B(2D)=1.28 \pm 0.02$. The range where the linear regression is fitted is almost two orders of magnitude. 


\begin{figure}
\begin{center}
\includegraphics[scale=1]{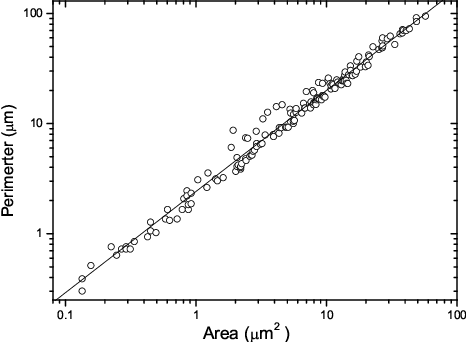}
\end{center} \caption{\label{fig:dp_example_2} Linear regression
for the calculation of $D_p$ corresponding to a
magneto-rheological fluid with concentration $\phi_{2D}=0.088$ and
$\lambda=1718$ after applying the magnetic field during $4500$ s.
A value of $D_{p}=1.84$ is obtained.}
\end{figure}

Other methods for obtaining fractal dimensions consist on using scaling relations for
other fractal quantities, that are obtained from collectivities of objects. We can consider various types of these fractal dimensions: the so-named one and two dimensional fractal dimensions, $D_{1}$ and $D_{2}$, which are obtained by means of $P \sim l_{f}^{\,D_1}$ and $A \sim l_{f}^{\,D_2}$, where $P$ is the perimeter of a cluster, $A$ the area and
$l_f$ the longest distance between two points in the cluster, the
so-called Feret's diameter \cite{Lee2004}. Another useful expression uses the area and the perimeter for obtaining the perimeter-based fractal dimension $D_p$ in the following
form:
\begin{equation}\label{Dpf}
A \sim P^{\,2/D_p}
\end{equation}
Thus, when the contour of the clusters or aggregates is detected by means of
image analysis, we only need to calculate the perimeter, the area
and the Feret's diameter for obtaining the corresponding 2D
fractal dimension. This calculation is applied to all the objects contained in every image. Fig. \ref{fig:dp_example_2} is a typical example of the data obtained in such a way. The power law behavior is verified in
three orders of magnitude. However, some critics have been lately
reported \cite{Imre2006} about the applicability of
the area-perimeter method. The main argument of that critic lies on
the fact that digitizing resolution can change the perimeter and area of
the objects of study, if it is applied for objects oriented in
different directions. In our case, this argument is not
applicable, because of the anisotropy of the clusters caused by
the external magnetic field.

We fit three linear regressions in
every captured image (every $0.4$ s) using our software, aiming
to observe the changes on these fractal dimensions during
aggregation and disaggregation. We require a regression
correlation factor of at least $r>0.98$ for considering that the
result of its respective image is correct. This condition 
limits the analysis at the beginning of the aggregation and at the end of the disaggregation processes. We use a wide field of view for
detecting as many chains as possible.

\subsubsection{Roughness.}

We also compute some magnitudes related to the roughness of the long aggregates, which will be useful in the interpretation of the physical meaning of the fractal dimensions obtained in this work. We are interested in determining the deformation on the contour line of the analyzed clusters. When the magnetic field is applied, the variations on the contour line should not be large. However, when the field is disconnected we observe a great variation on the contour of the objects that can be studied in term of roughness. Specifically, we compute the height of the clusters contour measured
from a central line which crosses the clusters end to end. We named the height of the clusters as $h_j(i,t)$, where $j$ refers to a given cluster. This quantity depends of the position $i$ of the contour point and time $t$. The average height of the cluster contour $j$ is $\overline{h_j(t)}=1/N_j\,\sum_{i=1}^{N_j}h_j(i,t)$, where $i$ refers to the corresponding contour point and $N_j$ is the total number of contour points corresponding to the cluster $j$. The contour roughness of chains formed by microparticles has been previously studied on fluctuating but stable chains \cite{Silva96, Toussaint}. However, in our case we are interested on the disaggregation of the chains by Brownian movement of the particles.

We also calculate the root mean square of the border height fluctuations for the cluster $j$ as:
\begin{equation}\label{fluctuation}
w_j^2(t)\equiv {\frac{1}{N_j}\sum_{i=1}^{N_j}\left[h_j(i,t)-\overline{h_j(t)}\right]^2}
\end{equation}
This quantity characterizes the roughness of the contour in a fixed time. If we want to calculate the
cluster average value of the border height fluctuation, we can use the expression $W(t)\equiv {\left<w_j(t)\right>}_j = 1/N_c\,\sum_{j=1}^{N_c}w_j(t)$, being $N_c$ the number of clusters in the image and where the mean is made for all the clusters contained in the image. We also use the temporal evolution of the average height-height correlation $W(t)$ for observing the changes on the cluster roughness during the aggregation and disaggregation processes.

We can also define the height-height correlation:
\begin{equation}\label{correlation}
c_j^2(l,t)=\left<\left[h_j(x)-h_j(x')\right]^2\right>
\end{equation}
where $l=|x-x'|$ is the distance between two contour points
projections, $x$ and $x'$, made into a central end to end line
that crosses the cluster. This magnitude scales in the form
$c_j(l)\sim l^{\alpha}$, being $\alpha$ the roughness exponent. This
property has been widely used, for instance, for the study of
surface growth \cite{Barabasi1995} and it has been shown to be more robust 
against resolution effects than other roughness related properties \cite{Buceta2000}.
Moreover, it is possible to connect the roughness exponent with a
fractal dimension $d$ \cite{Barabasi1995}:
\begin{equation}\label{dimalpha}
d=2-\alpha
\end{equation}
It is accepted that for self-affine fractals, the fractal dimension $d$ corresponds to the box-counting dimension $D_B(2D)$ \cite{Feder1988}. Nevertheless, we should take into account that the relation (\ref{dimalpha}) varies depending on the fractal dimension that we are using \cite{Balankin1997}. The Eq.(\ref{dimalpha}) should be only correct for the Hausdorff dimension or for the box-counting dimension using self-affine fractals.

\section{Results and Discussion}\label{results}

\subsection{Aggregation.}\label{aggregation}

\begin{figure}
\begin{center}
\includegraphics[scale=0.9]{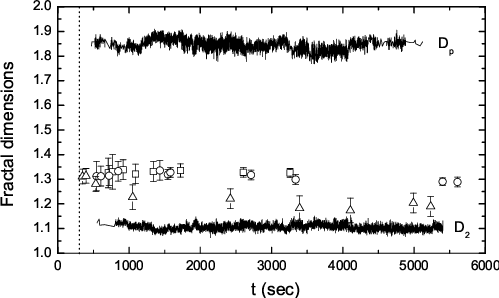}
\end{center} \caption{\label{fig:signal}Temporal evolution of
projected projected fractal dimensions during aggregation.
Continuous lines $D_2$ (bottom) and $D_p$ (top) for a case of
$\lambda = $812 and $\phi_{2D}=$0.106 (correlation factor
$r>0.98$). The single points are the box-counting dimension
$D_B(2D)$ for three different experiments: $\lambda=$77 and $\phi_{2D}=$0.071 for squares;
$\lambda=$171 and $\phi_{2D}=$0.115 for circles; $\lambda=$812 and
$\phi_{2D}=$0.106 for triangles.}
\end{figure}

We calculate the fractal dimensions explained in the previous sections for different experiments varying the external parameters $\lambda$ and $\phi_{2D}$. In Fig. \ref{fig:signal}, we display an example of the values of
$D_2$ and $D_p$ obtained along the aggregation process with the
requirement of $r>0.98$, reducing the number of useful images to $~$10000 images for experiment. In this figure, we also show the average values of $D_B(2D)$ for three different experiments. No temporal variation of
these fractal dimensions is observed, for each experiment and during aggregation. However, we can see that the values for $D_B(2D)$ are different depending on the experiment, i.e., they change when the external parameters $\lambda$ and $\phi_{2D}$ are varied.

Therefore, in the following we assume that making temporal averages of these fractal dimensions for each experiment is a correct procedure, because there is evolution of the distribution of cluster size for $t>$ 500 s \cite{Dominguez-Garcia2007} and due to the fact that no temporal dependency is observed. The obtained values for $D_2$ and $D_p$ are summarized on Table \ref{tab:table1} on columns 4 and 5, the
associated $\lambda$ and $\phi_{2D}$ values can be consulted on
columns 1 and 2. No dependency with $\lambda$, $\phi_{2D}$ or with
the ratio $R_1/R_0$ is observed on these two fractal dimensions. We also calculate the one
dimensional dimension $D_1$ for the chains, with no variation
or dependencies during aggregation.

\begin{center}
\begin{table}
\caption{\label{tab:table1}{2D fractal dimensions and $\beta$
exponents (Units for $\left<W\right>_t$ are $\mu m$.)}}
\vspace{0.5cm}
\setlength{\tabcolsep}{1.2pt}
\begin{tabular}{|ccc|cccccc|}
\hline $\lambda$ &$\phi_{2D}$ &$R_1/R_0$ &$D_2$ &$D_p$ &$D_B^{2D}$
&$\left<W\right>_t$
&$\beta_{d1}$ &$\beta_{d2}$\\
\hline\hline $77$ &$0.071$ &$1.28$ &$1.14$ &$1.85$ &$1.33$ &$0.21$
&$0.23$ &-$0.16$\\
$77$ &$0.132$ &$1.74$ &$1.03$ &$1.89$ &$1.28$ &$0.17$
&$0.19$ &-$0.11$\\
$171$ &$0.068$ &$1.64$ &$1.09$ &$1.89$ &$1.30$ &$0.20$
&$0.14$ &-$0.15$\\
$171$ &$0.115$ &$2.13$ &$1.10$ &$1.84$ &$1.31$ &$0.17$
&$0.17$ &-$0.10$\\
$296$ &$0.145$ &$2.87$ &$1.07$ &$1.84$ &$1.26$ &$0.16$
&$0.15$ &-$0.09$\\
$455$ &$0.112$ &$2.91$ &$1.1$ &$1.87$ &$1.27$ &$0.19$
&$0.17$ &-$0.09$\\
$640$ &$0.051$ &$2.19$ &$1.10$ &$1.82$ &$1.30$ &$0.22$
&$0.16$ &-$0.13$\\
$640$ &$0.086$ &$2.85$ &$1.05$ &$1.92$ &$1.27$ &$0.18$
&$0.20$ &-$0.12$\\
$812$ &$0.038$ &$2.06$ &$1.11$ &$1.84$ &$1.28$ &$0.21$
&$0.10$ &-$0.07$\\
$812$ &$0.106$ &$3.43$ &$1.10$ &$1.85$ &$1.20$ &0.18
&$0.15$ &-$0.13$\\
$985$ &$0.051$ &$2.54$ &$1.05$ &$1.87$ &$1.27$ &$0.25$
&$0.06$ &-$0.12$\\
$985$ &$0.075$ &$3.08$ &$1.05$ &$1.85$ &$1.25$ &$0.26$
&$0.19$ &-$0.08$\\
$1531$ &$0.059$ &$3.17$ &$1.08$ &$1.84$ &$1.25$ &$0.22$
&$0.03$ &-$0.12$\\
$1718$ &$0.088$ &$4.01$ &$1.10$ &$1.87$ &$1.22$ &$0.15$
&$0.21$ &-$0.09$\\
$1909$ &$0.045$ &$2.97$ &$1.09$ &$1.82$ &$1.22$ &$0.22$
&$0.21$ &-$0.15$\\
$2844$ &$0.043$ &$3.33$ &$1.07$ &$1.78$ &$1.25$ &$0.3$
&$0.07$ &-$0.14$\\
\hline
\end{tabular}
\end{table}
\end{center}

Results of averaging the 2D fractal dimensions are summarized in Table \ref{tab:table2}. For $\left<D_1\right>$ it is obtained an average value in good agreement with the Euclidean value $D_1=1$. Two-dimensional fractal dimensions $D_2$ and $D_p$ are more useful
than one-dimensional fractal dimension $D_1$ in terms of
characterizing the formed chains from our magneto-rheological
system. For two-dimensional fractal dimension $D_2$, it is obtained a value quite close to the expected $D_2 = 1$ for a linear object in an Euclidean geometry. This result is also compatible with the previous value obtained by Helgesen \emph{et al.}, when an external magnetic field of $H=1$ Oe is applied and linear rod-like chains are formed. For this experiment, they obtain $D_2 = 1.05 \pm 0.03$ with $K_{dd}=1360$, equivalent to $\lambda=680$, being a result perfectly compatible with ours because we observe no dependence of $D_2$ with $\lambda$.

\begin{table}
\caption{\label{tab:table2}{Average 2D fractal dimensions during aggregation.}}
\vspace{0.2cm}
\setlength{\tabcolsep}{7pt}
\begin{center}\begin{tabular}{|ccc|}
\hline $\left<D_1\right>$ &$\left<D_2\right>$ &$\left<D_p\right>$\\
\hline $1.01 \pm 0.03$ &$1.09 \pm 0.02$ &$1.84 \pm 0.02$\\
\hline
\end{tabular}\end{center}
\end{table}

For the perimeter-based dimension $D_p$, we obtain an average value of $\left<D_p\right> =
1.84 \pm 0.02$. Both values, $D_2$ and $D_p$, must be contained in a
range between 1 and 2. For $D_p=1$, we have a perfect spherical
object, whereas for $D_p=2$, we obtain a linear object. In our
case, we obtain a value close to 2, corresponding to linear
objects. The main practical difference among $D_2$ and $D_p$ is
that $D_p$ is obtained from magnitudes which are calculated, so that gives more information about the form and shape of the object (area and perimeter), whereas $D_2$ (and $D_1$) are calculated by means of Feret's diameter, that only tells about the longest distance between two points on the cluster contour, quantity that gives information that may be the same for very different cluster morphologies. Therefore, $D_p$ reflects
better the roughness of the boundary of the aggregate. For this
reason, a value for $D_p$ is obtained not so close to 2 than $D_2$
to 1.

\begin{figure}
\begin{center}
\includegraphics[scale=0.9]{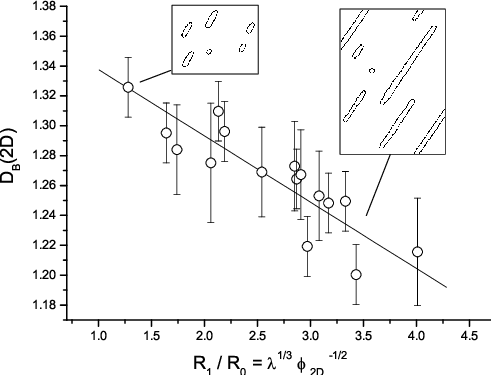}
\end{center} \caption{\label{db_agre2}Dependency of $D_B(2D)$ with
the ratio $R_1/R_0$ during aggregation (Table \ref{tab:table1},
columns 6 and 3). The continuous line is a linear
regression of the experimental data giving $D_B(2D)=$(-0.044 $\pm$
0.008)$R_1/R_0$ + (1.38 $\pm$ 0.02) with a correlation factor
$r=$-0.88.}
\end{figure}

In contrast to $D_1$, $D_2$ and $D_p$, the two-dimensional capacity dimension, $D_B(2D)$,
seems to show a relation with the external parameters, so an
average value for all the experiments cannot be calculated. In fact, we find a linear
dependence with the ratio $R_1 / R_0$, as it can be seen on Fig. \ref{db_agre2}. As we mentioned on the Introduction, the ratio
$R_1 / R_0$ informs about the aggregation being field driven
($R_1 > R_0$) or diffusion limited ($R_1 <
R_0$). In all our experiments, $R_1/R_0>1$ is verified, therefore,
from the beginning, the aggregation process is dominated by the
magnetic interaction among particles. Moreover, this result shows
that the shape of the chains depends on the intensity of this
ratio. Visually, we observe that the finer and longer formed
chains at the stationary state correspond to higher values of the
ratio $R_1/R_0$, while wider chains are observed when the ratio is
lower. In the figure, we show two snapshots of experiments with different values of $R_1/R_0$, as an illustration of this explanation. The capacity dimension measures how the chains fill the surrounding space, and therefore, Fig. \ref{db_agre2} shows the relationship between the shape of the clusters and the intensity of this magnetic interaction between particles. In our previous work on aggregation dynamics \cite{Dominguez-Garcia2007}, we show how relative difference between dynamical exponents also depends on this ratio. This result is another confirmation on the dependence of the MR fluid behavior with the amplitude of the external magnetic field and the concentration of particles.

\begin{figure}
\begin{center}
\includegraphics[scale=0.95]{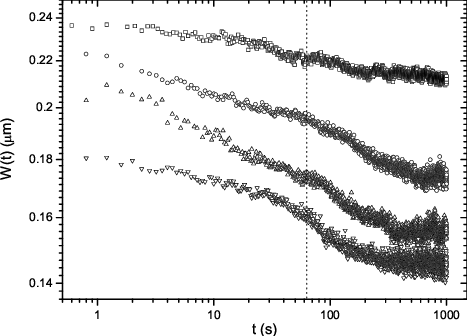}
\end{center} \caption{\label{fig:wagre}Four experiments as examples
of power-law behavior of the average chain-border height
fluctuation $W(t)$ in the initial stages of aggregation ($t<100$)
until saturation. From top to bottom: $\lambda=$77 and
$\phi_{2D}=$0.071; $\lambda=$812 and $\phi_{2D}=$0.106;
$\lambda=$296 and $\phi_{2D}=$0.145; $\lambda=$1718 and
$\phi_{2D}=$0.088.}
\end{figure}

For our experiments, we observe that
$W(t)$ is approximately constant during aggregation from $\sim 100$ s,
until the field is switched off (see Fig. \ref{fig:wagre}). The time of 100 seconds is approximately the time of formation of doublets of chains of only two particles in our experiments \cite{Dominguez-Garcia2007}. Therefore, this initial observed experimental decrease on the roughness is related to the behavior of the free particles. When the chains reach the mean size of the doublet, $W(t)$ remains constant and a average value can be calculated for each experiment. In
Table \ref{tab:table1}, column 7, we show time average values
$\left< W\right>_t$ for each experiment in the saturated region.
No dependence with $\lambda$, $\phi_{2D}$ or $R_1/R_0$ is
observed. Therefore, as it happens with $D_p$, the variations on
the border of the clusters do not suffer high variations during
aggregation and they seem not to depend on external variables. The
average value of this dispersion is 0.21$\pm$ 0.04 $\mu$m,
approximately $1/5$ of the particle diameter. This value can be
considered high for linear chains, because a perfect spherical
single particle should have a value of $w$, close to $1/3$ of the
particle diameter. We have to take into account that $\left< W\right>_t$
has been obtained making a cluster average in which the free
particles and small chains count the same as the longest clusters, something that decreases the value of $W(t)$ for a perfect chain
composed of aligned spherical particles.

\begin{figure}
\begin{center}
\includegraphics[scale=0.9]{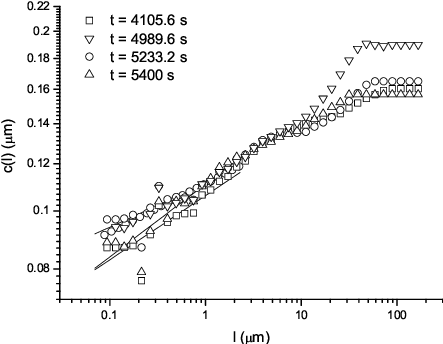}
\end{center}
\caption{\label{graph:correlacion}Height-height correlation function $c(l,t)$ for four chains in an experiment with $\lambda=812$ and $\phi_{2D}=0.106$ during aggregation: (rectangles) $t=4105.6 s$ s after the field is connected, we obtain a roughness exponent of $\alpha=0.11\pm 0.02$, (inverted triangles) $t=4989.6$ s and $\alpha=0.07 \pm 0.01$, (circles) $t=5233.2$ s and $\alpha=0.12 \pm 0.02$, (triangles) $t=5400$ s and $\alpha=0.08 \pm 0.01$. Inset: Example of chain contour image for $t=4989.6$ s.}
\end{figure}

In Fig. \ref{graph:correlacion}, we show an example of calculation of $c(l,t)$ for four chains with enough length (from 20 to 100 $\mu$m) for different times. We calculate a roughness exponent $\alpha$ on the interval $0.1\,\mu$m $< l <2 \,\mu$m, corresponding to experiment with $\lambda$=812 and $\phi_{2D}$ =
0.106, for different times: $t=$  4105.6 s, 4989.6 s, 5233.2 s and 5400 s, obtaining roughness exponents $\alpha=0.11\pm 0.02$, $\alpha=0.07 \pm 0.01$, $\alpha=0.12 \pm 0.02$ and $\alpha=0.08 \pm 0.01$, respectively. No substantial difference on this $\alpha$ values is observed; therefore, we make an average value of these chains 
$\left<\alpha\right>$ = 0.10. If we apply Eq.(\ref{dimalpha}), we calculate an average value of $\left<d\right>$ = 1.90, slightly larger but similar to the value $D_p \sim 1.85$ obtained for this experiment and with the average value $\left<D_p\right> = 1.84
\pm 0.02$. This approximate agreement between values calculated by means of $d=2-\alpha$ and $D_p$ is very surprising, because the perimeter-based dimension is not the adequate dimension for this equation. However, we have seen that the capacity dimension is related to how the clusters fill the space and they are not associated with the roughness or the contour of the objects. On the other hand, the magnitude that is more sensitive to the contour of the clusters is $D_p$, so it would be expected that this dimension shows information about the cluster contour, as $\alpha$ seems to do.

\subsection{Disaggregation.}\label{disaggregation}

\begin{figure}
\begin{center}
\includegraphics[scale=0.9]{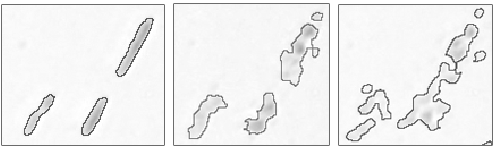}
\end{center}
\caption{\label{fig:desas}Example of chains disaggregating for
different times in the experiment $\lambda=77$ and
$\phi_{2D}=0.071$. The contour around the chains is the border of
the clusters detected by means of the image analysis, while the
grey color is the original captured image. Left: $t=0$ s (when
field is off), Middle: $t=20$ s, and Right: $t=40$ s}
\end{figure}

As stated above, because of our experimental image resolution, it is difficult to obtain reliable results on fractal dimensions when free particles regime dominate. That is why we cannot observe the variation on fractal dimensions while the chains are being formed by the aggregation of individual particles. However, we can observe the
disaggregation process, i.e, the separation by Brownian motion of the particles that formed the chains when the applied field is switched off. In Fig. \ref{fig:desas}, we show three snapshots of a group of chains during disaggregation at times $t=0$ (field is turned off), $t=20$ s and $t=40$ s. The dark border along the clusters border is the contour
extracted using our software. The images show how the free
particles have appeared at 20 seconds and how the clusters contour
folds onto itself. The snapshots shown in Fig. \ref{fig:desas} represent the regular behavior of the clusters when the field has been switched off.

\begin{figure}
\begin{center}
\includegraphics[scale=0.8]{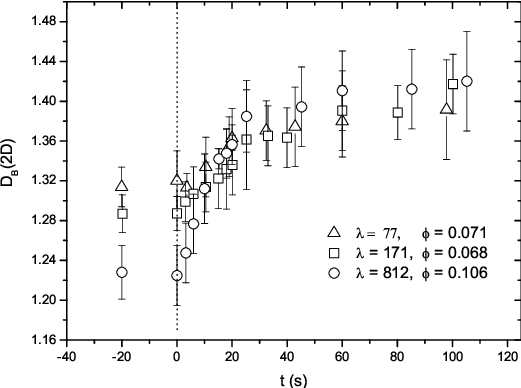}
\end{center}
\caption{\label{graph:db}Box-counting dimension during
disaggregation for three different experiments. The process begins
in each case with different $D_B(2D)$ values, but the fractal
dimension tends to a common temporal evolution whereas the chains
lose its linear structure.}
\end{figure}

We calculate some average values for the box-counting
dimension during disaggregation using the method explained in the Introduction. These calculations can be seen on
Fig. \ref{graph:db}. As we see on the study of aggregation, the
box-counting dimension is different for each experiment depending
on the ratio $R_1/R_0$; therefore, the experiments begin the
disaggregation process with different values of $D_B(2D)$.
However, when the field is switched off, all the values tend to a
single curve. This might be the expected behavior, if we suppose
that the capacity dimension measures how the clusters occupy the
space, since the clusters tend to dissolve into little clusters
and particles, regardless of the characteristics of the field and
concentration at beginning of the aggregation process.

As in the aggregation process, we can characterize the roughness of the chains when the field is disconnected. In Fig. \ref{fig:clt1}, 
we show the results of this kind of calculation using the height-height correlation function $c(l,t)$ for a 
chain in the process of disaggregation in an experiment with $\lambda=$812 and $\phi_{2D}=$0.106, in the 
range of $0.2-4.4$ seconds after the field is switched off. This limited interval can be explained because 
of the changing shape of the aggregates when the Brownian motion of the particle is predominant. 
For calculating the height-height correlation function, each $x$ point of the contour must be associated 
to only one height. As the disaggregation process advances and the free particles appear, the detected contour 
becomes not single-valued (see Fig. \ref{fig:desas}). When this occurs, it is not possible to further use this method of calculation. For this reason, we
can only calculate $\alpha$ values for $t<4.5$ s. This time can be considered as a 
characteristic diffusion time for this experimental system. Previous micro-rheological 
measurements \cite{Dominguez-Garcia2007} allowed us to determine that the diffusion 
coefficient for our system is $D = 0.23 \pm 0.01 \,\mu$m$^{2}$/s. Therefore, if we suppose 
that a center of single particle contained in a chain has to move one particle diameter 
for not being an aggregate, the corresponding diffusion time will be $t_d \sim (2a)^2/D \sim 1^2/0.23 = 4.3$ s. 
As the particles move out of the aggregates, the contour of the chains begins to be observed as a not 
univocal curve and the $c(l,t)$ calculation is not reliable any more.

\begin{figure}
\begin{center}
\includegraphics[scale=0.95]{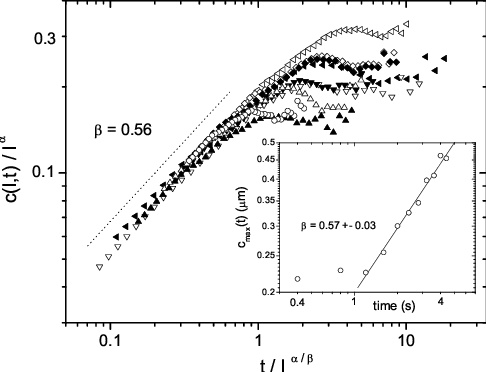}
\end{center} \caption{Family-Vicsek scaling function calculated by means of the height-height correlation function analysis of the roughness of the chains for the experiment with $\lambda=$812 and $\phi_{2D}=0.106$ shown on Fig. \ref{fig:clt1}. Inset: Time dependence of $c(l,t)$ saturated values from Fig. \ref{fig:clt1} for larger $l$. A growth exponent $\beta=0.57\pm0.03$ is obtained for $t>1$ s.} \label{fig:clt3}
\end{figure}

In Fig. \ref{fig:clt1} Inset, we show three images of aggregates for illustrating how their contour changes while the disaggregation process advances. The roughness exponent $\alpha$ for each chain is calculated on the interval $0.1\,\mu$m $< l <2 \,\mu$m as it is marked on Fig. \ref{fig:clt1} with dotted lines. Fig. \ref{fig:clt1} shows how the saturation values for $c(l,t)$ at larger $l$ are different depending on the time lapsed once the external field is disconnected. This result is shown in Fig. \ref{fig:clt3} Inset, where we plot the saturated value of $c(l,t)$ versus the corresponding time. Then, it is possible to obtain an exponent value $0.57 \pm 0.03$ for $t>1$ s, that we identify with the growth exponent $\beta$ \cite{Barabasi1995} that characterizes the time variation of the roughness of the chains. An interesting point is that the minimum value of the saturated height-height correlation function in Fig. \ref{fig:clt3} Inset, for $t=0.4$ s, is equal to 0.22 $\mu$m, which matches up with the average value during aggregation for the border height fluctuation $\left<W\right>_t=0.21 \pm 0.04$, previously related with the border height of a linear chain. Moreover, the maximum value obtained for saturated $c(l,t)$ for $t=4.4$ s is approximately equal to the radius of the particles $\sim 0.49\,\mu$m, showing that the particles begin to separate from the aggregates when that time is reached.

The height-height correlation function for this experiments should verify a Family-Vicsek scaling function \cite{Vicsek1984} such as:
\begin{equation}
 c(l,t)\sim l^{\alpha} f\left(\frac{t}{l^z}\right)\label{scaling}
\end{equation}
where $z$ is called the dynamic exponent and can be related with the other exponents by means of: 
\begin{equation}
z = \frac{\alpha}{\beta} \label{zeta}
\end{equation}
In Fig. \ref{fig:clt3}, we plot the scaling expression Eq.(\ref{scaling}) using Eq.(\ref{zeta}), the $\alpha$ exponent values calculated on Fig. \ref{fig:clt1} and the $\beta$ exponent calculated in Fig. \ref{fig:clt3} Inset, with $t>1$. In this figure we can show how all the data collapse, as expected, on a single curve showing that the Family-Vicsek scaling is verified for the roughness study of the aggregates. Moreover, the exponent calculated using this single curve for small values of $t/l^z$ provides a growth exponent $\beta=0.56 \pm 0.01$ in perfect agreement with the previously calculated one.

\begin{figure}
\begin{center}
\includegraphics[scale=0.9]{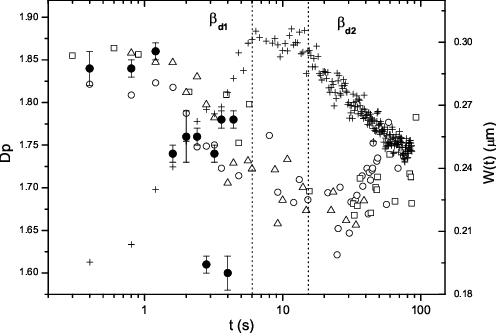}
\end{center} \caption{$D_p$ during disaggregation when the field is
switched off ($t=0$) in three different experiments (circles:
$\lambda=77$ and $\phi_{2D}=0.071$; squares: $\lambda=171$ and
$\phi_{2D}=0.068$ and triangles: $\lambda=$812 and
$\phi_{2D}=0.106$). The little crosses correspond to the average
chain-border height fluctuation $W(t)$. The black filled points
are $d=2-\alpha$ values with $\alpha$ obtained from Fig. \ref{fig:clt1}.} \label{fig:des}
\end{figure}

For the sake of completeness, we show in Fig. \ref{fig:des} the evolution of $D_p$ in the disaggregation process for three
different experiments varying $\lambda$ and $\phi_{2D}$ values (empty points). The perimeter-based fractal dimension $D_p$ changes from the value $D_p \sim 1.90$ for linear chains to a minimum value $D_p \sim 1.65$ after 20 seconds of
Brownian motion. After that, a stabilization on $D_p$ (for $t> 40$
s) is observed. Unfortunately this method has several
limitations. The first one is that the presence of free particles
reduces the correlation factor of the linear regression, making
the result less reliable. Secondly, this method may not be correct when the objects do
not have anisotropy, as occurs when the chain disaggregates
into free particles. Therefore, $D_p$ values may not be correct
when $t>60$ s during disaggregation.

We also show in Fig. \ref{fig:des} the
evolution of the average chain-border height fluctuation $W(t)$ (little crosses).
Two different states can be observed on this magnitude: the first is
a steep growth from $t=0$ to $t \sim 6$ s; the second is a soft
decrease of the curve from $t \sim 15$ s. This magnitude has,
two power-law separate behaviors in each region with different
exponents. We respectively named these exponents ${\beta}_{d1}$
and ${\beta} _{d2}$ and their values for each experiment have been
summarized on Table \ref{tab:table1}, columns 8 and 9. Average
calculations of both quantities give the following results:
$\left<{\beta}_{d1}\right>=0.15\pm 0.06$ and
$\left<{\beta}_{d2}\right>=-0.11\pm 0.03$. We interpret these two
regions as follows: the first region, the $\beta_{d1}$ region, is the one where
the roughness of the clusters grows due to the Brownian motion
of the particles to escape from the cluster. In this stage, the clusters
maintain their individuality not breaking yet into little pieces.
The second region, $\beta_{d2}$ region, appears when the clusters begin to break into little clusters and the free particles appear. At that point, the growth
on the average cluster roughness decreases, because the average is
calculated with the new clusters and free particles appear
during the process.

Finally, in the same figure, we show some values of $d=2-\alpha$, calculated using the roughness exponents of the scaling analysis (Fig. \ref{fig:clt1}). Then, in Fig. \ref{fig:des}, using black filled points, we plot $d=2-\alpha$ values for long chains in an experiment with $\lambda=$812 and $\phi_{2D}=0.106$, for $t<4.5$ s,  where the used methodology allows us to calculate the exponent $\alpha$. We can see how $d$ values fit adequately with the plotted $D_p$ points for the initial stages of disaggregation, but they seem to be different approximately to $\sim 4$ s. This initial agreement complies with the previously mentioned relationship between $D_p$ and $d$ during aggregation. The difference between these two quantities when we approximate to the diffusion time can be related with the different methods of calculation which particularly depend on the appearance of the free particles. 

\subsection{Extended discussion}\label{discussion}

Let us emphasize that different methods for the obtention of fractal dimension values actually take into account different geometrical properties of the system into consideration. Measures based on the Feret's diameter take into account only the statistical distribution of the cluster geometry considering all of the clusters present, but does not take into account the spatial distribution of the clusters. From a different point of view, measures related with chain roughness concern only the largest clusters present and the effects of thermal fluctuations on their form, but disregard small clusters and the spatial distribution of all of the clusters. Finally, the calculation of the box dimension upon the whole image would mix all the information concerning the geometry of the clusters themselves and the spatial distribution of the clusters, while the box dimension computed only on the largest clusters should take into account only the information on the local roughness of the chain as the roughness method does.

Hence, it is not surprising that the dimensions that do not take into account the spatial distribution of the chains are not sensitive to the control parameters of the experiment (for instance, $R_1/R_0$). Actually, the dimensions based on the Feret diameter ($D_1$ and $D_2$) or the perimeter area relationship ($D_p$) only consider, loosely speaking, the statistics of the chain size, which consists mainly of straight aligned objects with a width of one particles diameter. It is not surprising, therefore, that all of these dimensions yield values close to one, for $D_1$ and $D_2$, and close to two for $D_p$. The respective deviations from these integer values is due to the brownian fluctuations of the position of the particles in the direction perpendicular to the chain main axis.

On the other hand, the box dimension, $D_{B}(2D)$, takes into account the spatial distribution of the chains in the field of view too. This spatial distribution depends on the value of the control parameters. Actually for high fields and small particle density, i.e. high values of $R_1/R_0$, structures with few long chains and high inter-chain spacing appear, while at low fields and large particle density, i.e. low values of $R_1/R_0$, many short chains with small inter-chain spacing occur. This two types of limiting structures clearly need similar number of boxes when considering large box sizes. However, at small box sizes, a larger number of boxes is needed to cover the many short chains structure. Hence larger values of $D_{B}(2D)$ should be obtained at smaller values of $R_1/R_0$. This conjecture is confirmed by the results shown in Figure 6.       

From a different point of view, we would like to emphasize that the dependence of $D_{B}(2D)$ on the parameter $R_1/R_0$, shows that $D_{B}(2D)$ is a useful parameter to distinguish between field driven and diffusion driven regimes in aggregation processes of magneto-rheological fluids. Furthermore, the problem here studied is a paradigmatic case of a system in which different aggregation regimes (field driven or diffusion driven) can be achieved by changing the control parameters, with some resemblances to systems that change from diffusion driven to reaction driven \cite{Tirado2001}. Note, however, that in the later case \cite{Tirado2001} the diffusion driven regime is the fast one, while in the magneto-rheological fluid problem, the diffusion regime is the slow one. 

Now we turn to the results concerning the roughness of the long chains. Some understanding of the long chain contour dynamics can be gained by considering the chain model described in Ref. \cite{Toussaint}. If $h_i$ is the displacement of particle $i$ in the coordinate transversal to the main chain orientation, the equation that rules its evolution (in the inertialess approximation) is:
\begin{equation}
\label{eq:rouse}
\dot{h_i} = \frac{\alpha}{\kappa} (h_{i+1} + h_{i-1} -2h_i) + \frac{1}{\kappa} \xi_i(t),
\end{equation}
where $\alpha$ is a parameter directly proportional to the amplitude of the aligning magnetic field, $\kappa$ is proportional to the viscosity of the carrier fluid, $\kappa = 3\pi \eta  a$, and $\xi_i(t)$ is a fluctuating force due to the Brownian motion of the particles. Interestingly, for length scales larger that $a$, the continuum limit of the model coincides with the Rouse model of polymer chain dynamics \cite{Doi86,GrosbergKhokhlov}. 

The first stages of the disaggregation dynamics are conceivably ruled by the equation above with no external field, i.e., with $\alpha = 0$, which means that the evolution of the chain contour is ruled by the equation
\begin{equation}
\label{eq:rdm}
\dot{h_i} = \frac{1}{\kappa} \xi_i(t).
\end{equation}
In other words, the transversal displacements of the particles are independent random processes. Therefore, the system is equivalent to the so-called random deposition model in the surface growth literature \cite{Barabasi1995,Family1985}. Theoretical and numerical studies show that the evolution of the global surface roughness shows a scaling such as $W \sim t^{1/2}$, in good agreement with the value reported here. This $1/2$ time exponent appears also in the dynamics of Equation~\eqref{eq:rouse} at short times while the system evolves in a diffusion dominated regime \cite{Toussaint}.

However, no power law behavior with length scale is expected in the random deposition model, in clear contrast with the Family-Vicsek scaling obtained in the experiments. Hence, some type of interaction between particles is present that is being disregarded in the model. A possible candidate is hydrodynamic interaction. However,  the adequate model for chains with hydrodynamic interaction \cite{Toussaint} in an unbounded medium would be the Zimm model \cite{Doi86,GrosbergKhokhlov}, which predicts $W \sim t^{1/3}$. Exponents close to this value, namely $\beta = 3/8$, have been found in several magnetorheological systems \cite{Furst98,Furst00b,Cutillas,Silva96}.  However, in the case of chains located close to the cell walls hydrodynamic interactions should decay much faster than in unbounded media, so that a representation of the hydrodynamic interaction in terms of a long range force, such as in the Zimm model, should not be adequate for the dynamic of the chains reported here.

Long-range dipolar interactions have also been conjectured as responsible for the anomalous $\beta = 3/8$ time exponent \cite{Furst98,Furst00b,Toussaint}. However, we remind that in the experiments here reported, the field is switched off, so that magnetic dipolar interactions should not be present during the evolution of the chain contour. This point clearly deserves further investigation. 

\section{Summary.}\label{conclusion}

We study experimentally the morphology of aggregates of superparamagnetic micro-particles in water, i.e., a magneto-rheological fluid, when an external constant and uniaxial magnetic field is applied. We calculate 2D fractal dimensions and study the contour roughness of the clusters using image analysis. We have focused on two
processes: the aggregation, where the chains are formed by
the application of an external magnetic field, and the
disaggregation, which occurs when the magnetic field is switched
off. As far as we know, the disaggregation process has not been
studied in detail in the literature, in spite of its interest on
potential practical applications. In this work, we emphasize on the morphological interpretation of the calculated fractal dimension. We have also determined the region of application on all the magnitudes here used.

During aggregation, using the area-perimeter method, we obtain the following average values for: one-dimensional fractal dimension, $\left<D_1\right> = 1.01
\pm 0.03$; two-dimensional fractal dimension, $\left<D_2\right> = 1.09 \pm 0.02$ and perimeter-based fractal dimension, $\left<D_p\right> = 1.84 \pm 0.02$. We have compared these values with previous experimental works. The box-counting dimension ~---or capacity dimension $D_B(2D)$~--- does not vary with time, but its value is different
depending on the ratio $R_1/R_0$ between two characteristic lengths that measure a relative magnetic field strength per particle, reflecting that these fractal dimensions show how the chains occupy the surrounding space and agree with previous observed dependence on this ratio of kinetical exponents on this experimental system \cite{Dominguez-Garcia2007}. We have also calculated various quantities associated with the roughness of the contour or border of the clusters, particularly the height-height correlation function $c(l,t)$ that provides the roughness exponent $\alpha$.

A similar analysis has been developed for the data obtained during the process of chain disaggregation, i.e., when the applied field is switched off and the particles return to the initial situation of free particles. For box-counting dimension, the dependency with $R_1/R_0$ vanishes as the clusters dissolve. Moreover, we study $c(l,t)$ for time values lower than a characteristic diffusion time, and we obtain a time-dependence of the roughness variation. We calculate different coefficients such as the growth exponent $\beta = 0.57 \pm 0.03$ and the dynamic exponent $z$. These results allow us to verify that the Family-Vicsek scaling function, Eq.(\ref{scaling}), is verified for the disaggregating chains on the very first stage of the progress, when the particles are moving by Brownian motion inside the aggregate. Furthermore, we observe that $D_p$ decreases approximately as $2-\alpha$, and that the average chain-border height fluctuation $\left<W(t)\right>$ shows two different power-law behaviors during disaggregation.

\section{Acknowledgments.}
We wish to acknowledge J.M. Gonz\'alez, J.M.
Palomares and F. Pigazo (ICMM) for the VSM magnetometry
measurements, J.M. Pastor for fruitful discussions and J. C. G\'{o}mez-S\'{a}ez for her proofreading of the English texts. This research has been partially
supported by M.E.C. under Project No. FIS2006-12281-C02-02, and by
C.A.M under Project S/0505/MAT/0227.


\begin{thebibliography}{40}
\providecommand{\natexlab}[1]{#1}
\providecommand{\url}[1]{\texttt{#1}}
\expandafter\ifx\csname urlstyle\endcsname\relax
  \providecommand{\doi}[1]{doi: #1}\else
  \providecommand{\doi}{doi: \begingroup \urlstyle{rm}\Url}\fi

\bibitem[Rabinow(1948)]{Rabinow48}
J.~Rabinow.
\newblock The magnetic fluid clutch.
\newblock \emph{AIEE Trans.}, 67:\penalty0 1308, 1948.

\bibitem[Kerr(1990)]{Kerr}
R.~A. Kerr.
\newblock \emph{Science}, 247:\penalty0 050401, 1990.

\bibitem[{$N$}akano and Koyama(1998)]{Nakano98}
M.~{$N$}akano and K.~Koyama, editors.
\newblock \emph{Proceedings of the 6th International Conferences on ER and MR
  fluids and their applications}, 1998. World Scientific, Singapore.

\bibitem[Garc{\'i}a(2003)]{cortisol}
R.~Y. Garc{\'i}a.
\newblock \emph{Lock-in Signal Amplification Using Microrotors for Rapid and
  Sensitive Detection of Cortisol}.
\newblock PhD thesis, Arizona State University., May 2003.

\bibitem[Smirnov et~al.(2004)Smirnov, Gazeau, Lewin, Bacri, Siauve, Vayssettes,
  Cuenod, and Clement]{tumortherapy}
P.~Smirnov, F.~Gazeau, M.~Lewin, J.~C. Bacri, N.~Siauve, C.~Vayssettes, C.~A.
  Cuenod, and O.~Clement.
\newblock In vivo cellular imaging of magnetically labeled hybridomas in the
  spleen with a 1.5-t clinical mri system.
\newblock \emph{Magn. Reson. Medi.}, 52:\penalty0 73--79, 2004.

\bibitem[Helgesen et~al.(1988)Helgesen, Skjeltorp, Mors, Botet, and
  Jullien]{Helgesen1988}
G.~Helgesen, A.~T. Skjeltorp, P.~M. Mors, R.~Botet, and R.~Jullien.
\newblock Aggregation of magnetic microspheres: Experiments and simulations.
\newblock \emph{Phys. Rev. Lett.}, 61\penalty0 (15):\penalty0 1736--1739, 1988.

\bibitem[Mart\'{i}nez-Pedrero et~al.(2005)Mart\'{i}nez-Pedrero, Tirado-Miranda,
  Schmitt, and Callejas-Fern\'{a}ndez]{Martinez-Pedrero2005}
F.~Mart\'{i}nez-Pedrero, M.~Tirado-Miranda, A.~Schmitt, and
  J.~Callejas-Fern\'{a}ndez.
\newblock Aggregation of magnetic polystyrene particles: A light scattering
  study.
\newblock \emph{Colloids Surf., A}, 270-271:\penalty0 317--322, 2005.

\bibitem[Mart\'{i}nez-Pedrero et~al.(2006)Mart\'{i}nez-Pedrero, Tirado-Miranda,
  Schmitt, and Callejas-Fern\'{a}ndez]{Martinez-Pedrero2006}
F.~Mart\'{i}nez-Pedrero, M.~Tirado-Miranda, A.~Schmitt, and
  J.~Callejas-Fern\'{a}ndez.
\newblock Forming chainlike filamentes of magnetic colloids: The role of the
  relative strength of isotropic and anisotropic particle interactions.
\newblock \emph{J. Chem. Phys.}, 125:\penalty0 084706, 2006.

\bibitem[Mart\'{i}nez-Pedrero et~al.(2007)Mart\'{i}nez-Pedrero, Tirado-Miranda,
  Schmitt, Vereda, and Callejas-Fern\'{a}ndez]{Martinez-PedreroCSA2007}
F.~Mart\'{i}nez-Pedrero, M.~Tirado-Miranda, A.~Schmitt, Fernando Vereda, and
  J.~Callejas-Fern\'{a}ndez.
\newblock Structure and stability of aggregates formed by electrical
  double-layered magnetic particles.
\newblock \emph{Colloids Surf., A}, 306:\penalty0 158--165, 2007.

\bibitem[Vicsek(1992)]{viskecbook}
T.~Vicsek.
\newblock \emph{Fractal Growth Phenomena}.
\newblock World Scientific, Singapore, 2 edition, 1992.

\bibitem[Jullien and Botet(1987)]{jullienbook}
R.~Jullien and R.~Botet.
\newblock \emph{Aggregation and Fractal Aggregates}.
\newblock World Scientific, Singapore, 2 edition, 1987.

\bibitem[Ding and Liu(1989)]{Ding1989}
J.~R. Ding and B.~X. Liu.
\newblock Fractal aggregation of magnetic particles in {A}g-{C}o thin-film
  surfaces.
\newblock \emph{Phys. Rev. B}, 40\penalty0 (8):\penalty0 5834--5836, 1989.

\bibitem[Niklasson et~al.(1988)Niklasson, Torebring, Larsson, Granqvist, and
  Farestam]{Niklasson1988}
G.~A. Niklasson, A.~Torebring, C.~Larsson, C.~G. Granqvist, and T.~Farestam.
\newblock Fractal dimension of gas-evaporated {C}o aggregates: Role of magnetic
  coupling.
\newblock \emph{Phys. Rev. Lett.}, 60:\penalty0 1735--1738, 1988.

\bibitem[Carrillo et~al.(2003)Carrillo, Donado, and Mendoza]{Carrillo2003}
J.~L. Carrillo, F.~Donado, and M.~E. Mendoza.
\newblock Fractal patterns, cluster dynamics, and elastic properties of
  magnetorheological suspensions.
\newblock \emph{Phys. Rev. E}, 68:\penalty0 061509 (1--8), 2003.

\bibitem[Shen et~al.(2001)Shen, Stachowiak, Fateen, Laibinis, and
  Hatton]{Shen2001}
L.~Shen, A.~Stachowiak, S.~E.~K. Fateen, P.~E. Laibinis, and T.~A. Hatton.
\newblock Structure of alkanoic acid stabilized magnetic fluids. {A}
  small-angle neutron and light scattering analysis.
\newblock \emph{Langmuir}, 17:\penalty0 288, 2001.

\bibitem[Licinio and Fr\'{e}zard(2001)]{Licinio2001}
P.~Licinio and F.~Fr\'{e}zard.
\newblock Diffusion limited field induced aggregation of magnetic liposomes.
\newblock \emph{Braz. J. Phys.}, 31 (3):\penalty0 356--359, 2001.

\bibitem[Bushell et~al.(2002)Bushell, Yan, Woodfield, Raper, and
  Amal]{Bushell2002}
G.~C. Bushell, Y.~D. Yan, D.~Woodfield, J.~Raper, and R.~Amal.
\newblock On techniques for the measurement of the mass fractal dimension of
  aggregates.
\newblock \emph{Adv. Colloid Interface Sci.}, 95:\penalty0 1--50, 2002.

\bibitem[Crocker and Grier(1996)]{Crocker1996}
J.~C. Crocker and D.~G. Grier.
\newblock Methods of digital video microscopy for colloidal studies.
\newblock \emph{J. Colloid Interface Sci.}, 179:\penalty0 298--310, 1996.

\bibitem[Lee and Kramer(2004)]{Lee2004}
C.~Lee and T.~A. Kramer.
\newblock Prediction of three-dimensional fractal dimensions using the
  two-dimensional properties of fractal aggregates.
\newblock \emph{Adv. Colloid Interface Sci.}, 112:\penalty0 49--57, 2004.

\bibitem[Maggi and Winterwerp(2004)]{Maggi2004}
F.~Maggi and J.~C. Winterwerp.
\newblock Method for computing the three-dimensional capacity dimension from
  two-dimensional projections of fractal aggregates.
\newblock \emph{Phys. Rev. E}, 69\penalty0 (011405), 2004.

\bibitem[S{\'a}nchez et~al.(2005)S{\'a}nchez, Alfaro, and
  P{\'e}rez]{Sanchez2005}
N.~S{\'a}nchez, E.~J. Alfaro, and E.~P{\'e}rez.
\newblock The fractal dimension of projected clouds.
\newblock \emph{Astrophys. J.}, 625:\penalty0 849--856, 2005.

\bibitem[Dom\'{i}nguez-Garc\'{i}a et~al.(2007)Dom\'{i}nguez-Garc\'{i}a, Melle,
  Pastor, and Rubio]{Dominguez-Garcia2007}
P.~Dom\'{i}nguez-Garc\'{i}a, S.~Melle, J.~M. Pastor, and M.~A. Rubio.
\newblock Scaling in the aggregation dynamics of a magneto-rheological fluid.
\newblock \emph{Phys. Rev. E}, 76:\penalty0 051403, 2007.

\bibitem[Ima()]{ImageJ2}
~{U}.~{S}. {N}ational {I}nstitutes of {H}ealth, {B}ethesda, {M}aryland, {USA},
  http://rsb.info.nih.gov/ij/.

\bibitem[Smith et~al.(1996)Smith, Lange, and Marks]{Smith1996}
T.~G. Smith, G.~D. Lange, and W.~B. Marks.
\newblock Fractal methods and results in cellular morphology: dimensions,
  lacunarity and multifractals.
\newblock \emph{J. Neurosci. Meth.}, 69:\penalty0 123--136, 1996.

\bibitem[Halley et~al.(2004)Halley, Hartley, Kallimanis, Kunin, J.Lennon, and
  Sgardelis]{Halley2004}
J.~M. Halley, S.~Hartley, A.~S. Kallimanis, W.~E. Kunin, J.~J.Lennon, and S.~P.
  Sgardelis.
\newblock Uses and abuses of fractal methodology in ecology.
\newblock \emph{Ecolog. Lett.}, 7:\penalty0 254--271, 2004.

\bibitem[Imre(2006)]{Imre2006}
A.~R. Imre.
\newblock Artificial fractal dimension obtained by using perimeter-area
  relationship of digitalized images.
\newblock \emph{Appl. Math. Comp.}, 173:\penalty0 443, 2006.

\bibitem[Silva et~al.(1996)Silva, Bond, Plourabou{\' e}, and Wirtz]{Silva96}
A.~S. Silva, R.~Bond, F.~Plourabou{\' e}, and D.~Wirtz.
\newblock Fluctuation dynamics of a single magnetic chain.
\newblock \emph{Phys. Rev. E}, 54:\penalty0 5502, 1996.

\bibitem[Toussaint et~al.(2004)Toussaint, Helgesen, and Flekkoy]{Toussaint}
R.~Toussaint, G.~Helgesen, and E.~G. Flekkoy.
\newblock Dynamic roughening and fluctuations of dipolar chains.
\newblock \emph{Phys. Rev. Lett.}, 93 (10):\penalty0 108304, 2004.

\bibitem[Barab\'{a}si and Stanley(1995)]{Barabasi1995}
A.~L. Barab\'{a}si and H.~E. Stanley.
\newblock \emph{Fractal concepts in surface growth}.
\newblock Cambridge University Press, 2 edition, 1995.

\bibitem[Buceta et~al.(2000)Buceta, Pastor, Rubio, and delaRubia]{Buceta2000}
J.~Buceta, J.~Pastor, M.~A. Rubio, and F.~J. delaRubia.
\newblock Finite resolution effects in the analysis of the scaling behavior of
  rough surfaces.
\newblock \emph{Phys. Rev. E}, 61\penalty0 (5):\penalty0 6015--6018, May 2000.
\newblock \doi{10.1103/PhysRevE.61.6015}.

\bibitem[Feder(1988)]{Feder1988}
J.~Feder.
\newblock \emph{Fractals}.
\newblock Plenum press, New York, 1988.

\bibitem[Balankin(1997)]{Balankin1997}
A.~S. Balankin.
\newblock Physics of fracture and mechanics of self-affine cracks.
\newblock \emph{Engin.~Fracture~Mech.}, 57\penalty0 (2-3):\penalty0 135--203,
  1997.

\bibitem[Vicsek and Family(1984)]{Vicsek1984}
T.~Vicsek and F.~Family.
\newblock Dynamic scaling for aggregation of clusters.
\newblock \emph{Phys. Rev. Lett.}, 52,19:\penalty0 1669--1672, 1984.

\bibitem[Odriozola et~al.(2001)Odriozola, Tirado-Miranda, Schmitt, L\'{o}pez,
  Callejas-Fern\'{a}ndez, Mart\'{i}nez-Garc\'{i}a, and
  Hidalgo-\'{A}lvarez]{Tirado2001}
G.~Odriozola, M.~Tirado-Miranda, A.~Schmitt, F.~M. L\'{o}pez,
  J.~Callejas-Fern\'{a}ndez, R.~Mart\'{i}nez-Garc\'{i}a, and R.~F.
  Hidalgo-\'{A}lvarez.
\newblock A light scattering study of the transition region between diffusion-
  and reaction-limited cluster aggregation.
\newblock \emph{J. Colloid Interface Sci.}, 240:\penalty0 90--96, 2001.

\bibitem[Doi and Edwards(1986)]{Doi86}
M.~Doi and S.F. Edwards.
\newblock \emph{The Theory of Polymer Dynamics.}
\newblock Clarendon Press, Oxford, 1986.

\bibitem[Grosberg and Khokhlov(1994)]{GrosbergKhokhlov}
A.~Y. Grosberg and A.~R. Khokhlov.
\newblock \emph{Statistical Physics of Macromolecules}.
\newblock AIP Press, New York, 1994.

\bibitem[Family and Vicsek(1985)]{Family1985}
F.~Family and T.~Vicsek.
\newblock Scaling of the active zone in the eden process on percolation
  networks and the ballistic deposition model.
\newblock \emph{J. Phys. A}, 18:\penalty0 L75--L81, 1985.

\bibitem[Furst and Gast(1998)]{Furst98}
E.~M. Furst and A.~P. Gast.
\newblock Particle dynamics in magnetorheological suspensions using
  diffusing-wave spectroscopy.
\newblock \emph{Phys. Rev. E}, 58:\penalty0 3372, 1998.

\bibitem[Furst and Gast(2000)]{Furst00b}
E.~M. Furst and A.~P. Gast.
\newblock Dynamics and lateral interactions of dipolar chains.
\newblock \emph{Phys. Rev. E}, 62:\penalty0 6916, 2000.

\bibitem[Cutillas and Liu(2001)]{Cutillas}
S.~Cutillas and J.~Liu.
\newblock Experimental study on the fluctuations of dipolar chains.
\newblock \emph{Phys. Rev. E}, 64:\penalty0 011506, 2001.

\end{thebibliography}

\end{document}